\documentclass[12pt]{iopart}
\usepackage{iopams}

\begin{document}
\title[Reduction of the Laplace sequence and sine-Gordon type equations]{Reduction of the Laplace sequence and sine-Gordon type equations}

\author{K I Faizulina$^{1}$, A R Khakimova$^{1}$}

\address{$^1$Institute of Mathematics, Ufa Federal Research Centre, Russian Academy of Sciences,
112, Chernyshevsky Street, Ufa 450008, Russian Federation}
\eads{\mailto{cherkira@mail.ru}, \mailto{aigul.khakimova@mail.ru}}

\begin{abstract}
In this work, we continue the development of methods for constructing Lax pairs and recursion operators for nonlinear integrable hyperbolic equations of soliton type, previously proposed in the work of Habibullin et al. (2016 {\it J. Phys. A: Math. Theor.} {\bf 57} 015203). This approach is based on the use of the well-known theory of Laplace transforms. The article completes the proof that for any known integrable equation of sine-Gordon type, the sequence of Laplace transforms associated with its linearization admits a third-order finite-field reduction. It is shown that the found reductions are closely related to the Lax pair and recursion operators for both characteristic directions of the given hyperbolic equation. Previously unknown Lax pairs and recursion operators were constructed.
\end{abstract}

\maketitle

\eqnobysec

\section{Introduction}

Nonlinear integrable equations are known for their applications in various fields of natural science, for example, in nonlinear optics \cite{NewellMoloney, Boyd, Gordon}, in hydrodynamics \cite{KdV, Hirota, Ablowitz, Zakharov86}, in acoustics \cite{Beyer}, in mathematical physics \cite{ZMNP, AblowitzClarkson}, in plasma physics \cite{Chen}, in condensed matter physics \cite{Shamsutdinov, Kittel}, in computer technology \cite{Morton}, in biology \cite{Nagumo, Murray} and etc. 

In this article we will focus on discussing nonlinear integrable hyperbolic equations of soliton type of the form
\begin{equation}\label{Iy.1}
u_{xy}=F(u,u_x,u_y).
\end{equation}
One of the brightest representatives of integrable equations of this type is the sine-Gordon equation
\begin{equation}\label{sG}
u_{xy}=\sin (u).
\end{equation}
It is widely used to simulate wave processes, propagation of solitons and other nonlinear wave structures. In particular, it describes the propagation of dislocations in crystals, waves in lipide membranes, and magnetic fields in Josephson transitions (see \cite{Zakharov86} and references). 

Due to their applications in various fields of science, nonlinear integrable equations are useful and interesting objects. As is known, for a detailed study and construction of particular solutions of these equations it is convenient to use such objects as the Lax pair, classical and higher symmetries, conservation laws, Dubrovin type equations, etc. In our recent papers \cite{HabKhaPo, HabKhFai}, we showed that the Laplace sequence provides an efficient way to construct Lax pairs of nonlinear hyperbolic equations. Moreover, at the same time, we also obtain recursion operators that describe hierarchies of higher symmetries for both characteristic directions of $x$ and $y$ of the hyperbolic type equation. It should be noted that along with the Lax representation and the recursion operator, one can also construct Dubrovin type equations (see, for example, \cite{HabKhaJPA20, UMJ21, Non23}). Let us note the work \cite{DubrovinMatveevNovikov}, in which the Dubrovin equations were derived using the method of finite-gap integration and algebraic-geometric solutions to the KdV equation were constructed.

Let us recall the main points of the cascade method of Laplace integration. Laplace cascade method \cite{Laplace} is known as one of the classical methods for constructing a general solution to a linear hyperbolic equation of the form
\begin{equation}\label{eq0}
u_{xy}+a(x,y)u_x+b(x,y)u_y+c(x,y)u=0.
\end{equation}
The main idea of this method is related to the sequence of Laplace invariants
\begin{equation}\label{LaplaceSeq}
\ldots, \, h_{[-2]}, \, h_{[-1]}, \, h_{[0]}, \, h_{[1]}, \, h_{[2]}, \, \ldots,
\end{equation}
the coefficients of which are calculated using the recurrent formula
\begin{equation*}
h_{[i+1]}=2h_{[i]}-h_{[i-1]}-(\ln (h_{[i]}))_{xy}, \quad i\in Z,
\end{equation*}
where $h_{[-1]}=b_y+ab-c$, $h_{[0]}=a_x+ab-c$. 
If the sequence of Laplace invariants (\ref{LaplaceSeq}) is terminated by zero at least on one side, i.e. if $h_{[k]}=0$ or $h_{[-k]}=0$, then the equation (\ref{eq0}) is integrated in quadratures. If the sequence of Laplace invariants is terminated by zero on both sides, then it is possible to construct a general solution without quadratures of the given equation.

Later, Laplace cascade method was adapted to the case of nonlinear integrable hyperbolic equations of Liouville type (see, for example, \cite{Liouville, Anderson, Dynnikov, Ferapontov, Adler, Zhiber, Ganzha}). 
In the work \cite{Zhiber} it was proven that if the sequence of Laplace invariants of the linearization of a nonlinear hyperbolic equation is finite, then the given nonlinear equation is Darboux integrable. Moreover, in this work it was shown that the Laplace cascade method can be used to classify nonlinear equations.

However, for a long time this approach did not find any interpretation in solitons theory. In the work \cite{HabKhaPo} a close connection between the theory of soliton equations and the theory of Laplace transforms was discovered. Research on this topic was continued in work \cite{HabKhFai} and this article. Our approach is based on the following hypothesis. We assume that for any integrable equation of the form (\ref{Iy.1}) that does not have nontrivial characteristic integrals, the sequence of Laplace transforms associated with its linearization (or Fr\'{e}chet derivative)
\begin{equation}\label{Iy.2}
v_{xy}+av_{x}+bv_{y}+cv=0
\end{equation}
where
\begin{equation}\label{abc}
a=-\frac{\partial F}{\partial u_x}, \quad b=-\frac{\partial F}{\partial u_y}, \quad c=-\frac{\partial F}{\partial u},
\end{equation}
admits a finite-field reduction, i.e. is compatible with some termination condition of the form 
\begin{eqnarray}\label{reduction}
\eqalign{
v_{[m+1]}=\alpha_{[-k]}v_{[-k]}+\alpha_{[-k+1]}v_{[-k+1]}+\cdots+\alpha_{ [m]}v_{[m]},\\
v_{[-k-1]}=\beta_{[-k]}v_{[-k]}+\beta_{[-k+1]}v_{[-k+1]}+\cdots+\beta_ {[m]}v_{[m]},}
\end{eqnarray}
where $m$, $k$ are integers such that $m\geqslant1$, $k\geqslant1$; the coefficients $\alpha_j, \beta_j$ depend on the dynamic variables $u$, $u_x$, $u_y, \ldots$. In addition, it is required that $\alpha_{[-k]}$ and $\beta_{[m]}$ do not vanish identically. Then the system of equations (\ref{reduction}) forms a Lax pair for the equation (\ref{reduction}), which differs from the usual Lax representation. From which, by means of elementary transformations, one can derive recursion operators for both characteristic directions and the usual Lax pair. In other words, these reductions are the foundation for constructing the recursion operator and Lax pair for nonlinear hyperbolic equations.

In (\ref{reduction}), the unknown functions $v_{[i]}$ are determined by iteratively applying the Laplace transform to the linearized equation (\ref{Iy.2}) as follows:
\begin{equation}\label{I.13-v}
\left(\frac{\partial}{\partial y}+a_{[i]}\right)v_{[i]}=v_{[i+1]}, \qquad \left(\frac{\partial}{\partial x}+b_{[0]}\right)v_{[i+1]}=h_{[i]}v_{[i]}
\end{equation}
and
\begin{equation}\label{I.14-v}
\fl \qquad \qquad \left(\frac{\partial}{\partial x}+b_{[-i]}\right)v_{[-i]}=v_{[-i-1]}, \qquad \left(\frac{\partial}{\partial y}+a_{[0]}\right)v_{[-i-1]}=k_{[-i]}v_{[-i]}.
\end{equation}
Here the coefficients $a_{[0]}:=a$, $b_{[0]}:=b$, $c_{[0]}:=c$ have the form (\ref{abc}), and $a_{[i+1]}$, $b_{[-i-1]}$, $h_{[i]}$ and $h_{[-i-1]}:=k_{[-i]}$, $i\geq 0$, are found according to the formulas (for more details see \cite{HabKhFai}):
\begin{eqnarray}
\fl \qquad h_{[0]}=D_x(a_{[0]})+a_{[0]}b_{[0]}-c_{[0]}, \quad & k_{[0]}=D_y(b_{[0]})+a_{[0]}b_{[0]}-c_{[0]}, \label{h0k0} \\
\fl \qquad a_{[i+1]}=a_{[i]}-D_y(\ln h_{[i]}), \quad & h_{[i+1]}=h_{[i]}+D_x(a_{[i+1]})-D_y(b_{[0]}), \label{ah+}\\
\fl \qquad b_{[-i-1]}=b_{[-i]}-D_x(\ln h_{[-i-1]}), \quad & h_{[-i-2]}=h_{[-i-1]}-D_x(a_{[0]})+D_y(b_{[-i-1]}).  \label{bh-}
\end{eqnarray}

In articles \cite{HabKhaPo, HabKhFai} and this work, we show that all known sine-Gordon type equations (see \cite{SokMesh}), namely equations (\ref{sG}) and
\begin{eqnarray}
 u_{xy}=f(u)\sqrt{1+u^2_x}, \quad f''=\gamma f, \quad \gamma=const; \label{2eq}\\
 u_{xy}=\sqrt{u_x}\sqrt{u_y^2+1}; \label{3eq} \\
 u_{xy}=\sqrt{u_x^2+1}\sqrt{u_y^2+1}; \label{4eq}\\
 u_{xy}=\sqrt{\wp(u)-\mu}\sqrt{u_x^2+1}\sqrt{u_y^2+q}, \label{5eq}
\end{eqnarray}
where $\wp$ is the Weierstrass function: $(\wp')^2=4\wp^3-g_2\wp-g_3$, $q$, $g_2$, $g_3$ are arbitrary constants, $\mu$ is a root of equation $4\mu^3-g_2\mu-g_3=0$, 
admit reductions of the form (\ref{reduction}) for $m=k=1$. Using the found reductions, Lax pairs and recursion operators for all sine-Gordon type equations were constructed. Note that for the equations (\ref{2eq})-(\ref{5eq}) these objects were not previously known.

{\bf  Remark 1.} In work \cite{HabKhFai} Lax pairs for equations (\ref{3eq}) and (\ref{4eq}) were found. There was a misprint in the Lax pair for the equation (\ref{3eq}), here we present its correct form:
\begin{equation*}
\left \{
\begin{array}{l}
  \displaystyle
\varphi_x=\xi\varphi+\frac{1}{2}\sqrt{u_x}\psi, \\[1ex]
  \displaystyle
\psi_x=\frac{1}{2}\sqrt{u_x}\varphi-\xi\psi,  
\end{array} 
\right.
\qquad 
\left \{
\begin{array}{l}
  \displaystyle
\varphi_y=\frac{1}{8}\xi^{-1}\left(u_y\varphi-\sqrt{u_y^2+1}\psi\right), \\[1ex]
  \displaystyle
\psi_y=\frac{1}{8}\xi^{-1}\left(\sqrt{u_y^2+1}\varphi-u_y\psi\right),  
\end{array} 
\right.
\end{equation*}
where $\xi=\frac{1}{4\sqrt{\lambda}}$.

This article is devoted to the study of equation (\ref{5eq}). Namely, in \S 2 we prove that this equation admits a reduction of the form (\ref{reduction}) for $m=k=1$ and present a third-order Lax pair, which differs from the usual one. In the same section, from the constructed reduction we derive recursion operators that describe hierarchies of higher symmetries for both characteristic directions of $x$ and $y$. Section 3 provides a detailed derivation of the Lax representation of the usual form. In the fourth section we discuss the equation (\ref{5eq}) in a more general form (see (\ref{gen-case})). For this equation we present the recursion operators and the Lax pair.

\section{Reductions of the Laplace sequence and recursion operators for equation (\ref{1.1(5)})}

The purpose of this section is to find a reduction of the Laplace sequence for an equation of sine-Gordon type
\begin{equation}\label{1.1(5)}
u_{xy}=\sqrt{\wp(u)-\mu}\sqrt{u_x^2+1}\sqrt{u_y^2+q}
\end{equation}
and to construct recursion operators that describe infinite hierarchies of higher symmetries for both characteristic directions for equation (\ref{1.1(5)}) .

Linear equation 
\begin{equation}\label{lin5}
v_{xy}+av_x+bv_y+cv=0
\end{equation}
is a linearization of equation (\ref{1.1(5)}), where the coefficients of the equation are given by formulas 
\begin{equation}\label{c1}
\eqalign{&a=-\frac{u_x\sqrt{\wp(u)-\mu}\sqrt{u_y^2+q}}{\sqrt{u_x^2+1}}, \quad b=-\frac{u_y\sqrt{\wp(u)-\mu}\sqrt{u_x^2+1}}{\sqrt{u_y^2+q}}, \cr\quad &c=-\frac{\wp'(u)\sqrt{u_x^2+1}\sqrt{u_y^2+q}}{2\sqrt{\wp(u)-\mu}}.}
\end{equation}
Consider a finite subsystem from an infinite system of equations (\ref{I.13-v}), (\ref{I.14-v}), which is obtained by repeatedly applying the Laplace transform to the linearized equation (\ref{lin5}):
\begin{equation}\label{subsys-1}
\eqalign{
\left(\frac{\partial}{\partial y}+a_{[1]}\right)v_{[1]}=v_{[2]}, \qquad \quad &\left(\frac{\partial}{\partial x}+b_{[0]}\right)v_{[2]}=h_{[1]}v_{[1]},\cr
\left(\frac{\partial}{\partial y}+a_{[0]}\right)v_{[0]}=v_{[1]}, \qquad &\left(\frac{\partial}{\partial x}+b_{[0]}\right)v_{[1]}=h_{[0]}v_{[0]},\cr
\left(\frac{\partial}{\partial x}+b_{[0]}\right)v_{[0]}=v_{[-1]}, \qquad &\left(\frac{\partial}{\partial y}+a_{[0]}\right)v_{[-1]}=k_{[0]}v_{[0]},\cr
\left(\frac{\partial}{\partial x}+b_{[-1]}\right)v_{[-1]}=v_{[-2]}, \qquad &\left(\frac{\partial}{\partial y}+a_{[0]}\right)v_{[-2]}=k_{[-1]}v_{[-1]}.
}
\end{equation}
Here the coefficients $a_{[0]}$, $b_{[0]}$ have the form (\ref{c1}), and the coefficients $h_{[0]}$, $k_{[0]} $, $ a_{[1]}$, $b_{[-1]}$, $h_{[1]}$, $k_{[-1]}$ are found using the formulas (\ref{h0k0}), (\ref {ah+}) and (\ref{bh-}).

{\bf Theorem 1.} 
The system of hyperbolic equations (\ref{subsys-1}) corresponding to the equation (\ref{1.1(5)}) admits reduction
\begin{equation}\label{1.9}
\eqalign{
v_{[2]}=\alpha_{[-1]}v_{[-1]}+\alpha_{[0]}v_{[0]}+\alpha_{[1]}v_{[1]},\cr
v_{[-2]}=\beta_{[-1]}v_{[-1]}+\beta_{[0]}v_{[0]}+\beta_{[1]}v_{[1]},}
\end{equation}
if the coefficients $\alpha_{[i]}$, $\beta_{[i]}$, $i=-1,0,1$, have the form
\begin{equation}\label{1.10}
\eqalign{
\alpha_{[-1]}=\frac{\lambda u_y\sqrt{u_y^2+q}}{\sqrt{\wp(u)-\mu}\sqrt{u_x^2+1}}, \quad \alpha_{[0]}=-q\lambda,\cr  \alpha_{[1]}=\frac{K(u_x^2+1)u_y}{-4(\wp(u)-\mu)u_{xx}+2\wp'(u)(u_x^2+1)}, \cr
\beta_{[-1]}=\frac{K(u_y^2+q)u_x}{-4(\wp(u)-\mu)u_{yy}+2\wp'(u)(u_y^2+q)}, \cr  \beta_{[0]}=-\frac{K}{4\lambda}, \quad \beta_{[1]}=\frac{Ku_x\sqrt{u_x^2+1}}{4\lambda\sqrt{\wp(u)-\mu}\sqrt{u_y^2+q}},}
\end{equation}
here $K=\frac{g_3}{\mu}+8\mu^2$ and $\lambda$ is a constant parameter.

{\bf Proof.}
In order to prove Theorem 1, we use equalities (\ref{subsys-1}) and (\ref{1.9}). Namely, to determine the unknown functions $\alpha_{[i]}$, $\beta_{[i]}$, we substitute (\ref{1.9}) into (\ref{subsys-1}), then we obtain the following equations:
\begin{equation}\label{1.11}
\eqalign{
v_{[1]y}=v_{[2]}-a_{[1]}v_{[1]}=(\alpha_{[1]}-a_{[1]})v_{[1]}+\alpha_{[0]}v_{[0]}+\alpha_{[-1]}v_{[-1]},\cr
v_{[0]y}=v_{[1]}-a_{[0]}v_{[0]},\cr
v_{[-1]y}=h_{[-1]}v_{[0]}-a_{[0]}v_{[-1]}}
\end{equation}
and
\begin{equation}\label{1.12}
\eqalign{
v_{[1]x}=h_{[0]}v_{[0]}-b_{[0]}v_{[1]},\cr
v_{[0]x}=v_{[-1]}-b_{[0]}v_{[0]},\cr
v_{[-1]x}=v_{[-2]}-b_{[-1]}v_{[-1]}=(\beta_{[-1]}-b_{[-1]})v_{[-1]}+\beta_{[0]}v_{[0]}+\beta_{[1]}v_{[1]}.}
\end{equation}
The remaining two equations, which we do not present, do not provide additional conditions for determining unknown functions.

From the compatibility condition $(v_{[i]x})_y=(v_{[i]y})_x$ of equalities (\ref{1.11}) and (\ref{1.12}), we derive a system of nonlinear equations for finding the required coefficients $\alpha_{[i]}$ and $\beta_{[i]}$:
\begin{equation}\label{1.13}
\eqalign{
D_y(\beta_{[-1]})+\beta_{[1]}\alpha_{[-1]}-h_{[-2]}=0,\cr
D_y(\beta_{[0]})+\beta_{[1]}\alpha_{[0]}+\beta_{[-1]}h_{[-1]}=0,\cr
D_y(\beta_{[1]})+\beta_{[1]}(\alpha_{[1]}+a_{[0]}-a_{[1]})+\beta_{[0]}=0,\cr
D_x(\alpha_{[-1]})+\alpha_{[-1]}(\beta_{[-1]}+b_{[0]}-b_{[-1]})+\alpha_{[0]}=0,\cr
D_x(\alpha_{[0]})+\alpha_{[-1]}\beta_{[0]}+\alpha_{[1]}h_{[0]}=0,\cr
D_x(\alpha_{[1]})+\alpha_{[-1]}\beta_{[1]}-h_{[1]}=0.}
\end{equation}
Here the coefficients $h_{[0]}$, $ h_{[1]}$, $ h_{[-1]}$, $ h_{[-2]}$, $a_{[0]}$, $ a_{[1]}$, $b_{[0]}$, $b_{[-1]}$ have the form
\begin{equation}\label{1.14}
\eqalign{ h_{[0]}=-\frac{\sqrt{\wp(u)-\mu}\sqrt{u_y^2+q}u_{xx}}{(u_x^2+1)^{3/2}}+\frac{\wp'(u)\sqrt{u_y^2+q}}{2\sqrt{\wp(u)-\mu}\sqrt{u_x^2+1}},\cr 
 h_{[1]}=\frac{Ku_xu_y(-4(\wp(u)-\mu)u_{xx}^2+(G-2\wp''(u))(u_x^2+1)^2)}{2\sqrt{\wp(u)-\mu}(-4(\wp(u)-\mu)u_{xx}+2\wp'(u)(u_x^2+1))}\cr
 \qquad +\frac{K(u_x^2+1)^{3/2}\sqrt{u_y^2+q}\sqrt{\wp(u)-\mu}}{-4(\wp(u)-\mu)u_{xx}+2\wp'(u)(u_x^2+1)}\cr
 \qquad +\frac{K(u_x^2+1)u_y}{(-2\sqrt{\wp(u)-\mu}u_{xx}+\sqrt{G}(u_x^2+1))^2}u_{xxx},\cr
 h_{[-1]}=-\frac{q\sqrt{\wp(u)-\mu}\sqrt{u_x^2+1}u_{yy}}{(u_y^2+q)^{3/2}}+\frac{q\wp'(u)\sqrt{u_x^2+1}}{2\sqrt{\wp(u)-\mu}\sqrt{u_y^2+q}},\cr
 h_{[-2]}=\frac{Ku_xu_y(-4(\wp(u)-\mu)u_{yy}^2+(G-2\wp''(u))(u_y^2+q)^2)}{2\sqrt{\wp(u)-\mu}(-4(\wp(u)-\mu)u_{yy}+2\wp'(u)(u_y^2+q))}\cr
 \qquad +\frac{K(u_y^2+q)^{3/2}\sqrt{u_x^2+1}\sqrt{\wp(u)-\mu}}{(-4(\wp(u)-\mu)u_{yy}+2\wp'(u)(u_y^2+q))}\cr
 \qquad +\frac{K(u_y^2+q)u_x}{(-2\sqrt{\wp(u)-\mu}u_{yy}+\sqrt{G}(u_y^2+q))^2}u_{yyy},}
\end{equation}
\begin{equation}\label{1.14-2}
\eqalign{
 a_{[0]}=-\frac{u_x\sqrt{\wp(u)-\mu}\sqrt{u_y^2+q}}{\sqrt{u_x^2+1}},\cr
 a_{[1]}=-\frac{u_yu_{yy}}{u_y^2+q}+\frac{u_x\sqrt{\wp(u)-\mu}\sqrt{u_y^2+q}}{\sqrt{u_x^2+1}}
-\frac{\wp'(u)u_y}{2(\wp(u)-\mu)}\cr
 \qquad +\frac{u_yK(u_x^2+1)}{-4(\wp(u)-\mu)u_{xx}+2\wp'(u)(u_x^2+1)},\cr
 b_{[0]}=-\frac{u_y\sqrt{\wp(u)-\mu}\sqrt{u_x^2+1}}{\sqrt{u_y^2+q}},\cr
 b_{[-1]}=-\frac{u_xu_{xx}}{u_x^2+1}+\frac{u_y\sqrt{\wp(u)-\mu}\sqrt{u_x^2+1}}{\sqrt{u_y^2+q}}
-\frac{\wp'(u)u_x}{2(\wp(u)-\mu)}\cr
 \qquad +\frac{u_xK(u_y^2+q)}{-4(\wp(u)-\mu)u_{yy}+2\wp'(u)(u_y^2+q)},}
\end{equation}
where $G=4\wp^2(u)+4\wp(u)\mu+\frac{g_3}{\mu}$.

Analyzing the equations (\ref{1.13}) and formulas for the coefficients (\ref{1.14})-(\ref{1.14-2}), we obtain that the functions $\alpha_{[i]}$, $\beta_{[i]}$ for $i=- 1,0,1$ can depend on dynamic variables $u_x$, $u_y$, $u_{xx}$, $u_{yy}$ as follows
\begin{equation*}
\eqalign{
\alpha_{[1]}=\alpha_{[1]}(u,u_x, u_y, u_{xx}), \qquad
&\beta_{[1]}=\beta_{[1]}(u,u_x, u_y),\cr
\alpha_{[0]}=\alpha_{[0]}(u,u_x, u_y), 
\qquad &\beta_{[0]}=\beta_{[0]}(u,u_x, u_y),\cr
\alpha_{[-1]}=\alpha_{[-1]}(u,u_x, u_y), 
&\beta_{[-1]}=\beta_{[-1]}(u,u_x, u_y,u_{yy}).}
\end{equation*}
Let us rewrite the system of equations (\ref{1.13}) taking into account the found representations of the sought functions $\alpha_{[i]}$ and $\beta_{[i]}$ and collect the coefficients for the independent variables (for the highest derivatives of the variable $u$ with respect to $x$ and $y$, on which the functions $\alpha_{[i]}$ and $\beta_{[i]}$ do not depend). We will not give an explicit form of the equations themselves because of their cumbersomeness.
In the first and last equations, we collect the coefficients of $u_{xxx}$ and $u_{yyy}$ and obtain equations for the unknown functions $\alpha_{[1]}$, $\beta_{[-1 ]}$
\begin{eqnarray*}
\alpha_{[1]u_{xx}}-\frac{K(u_x^2+1)u_y}{\left(-2\sqrt{\wp(u)-\mu}u_{xx}+\sqrt{G}(u_x^2+1)\right)^2}=0,\cr
\beta_{[-1]u_{yy}}-\frac{K(u_y^2+q)u_x}{\left(-2\sqrt{\wp(u)-\mu}u_{yy}+\sqrt{G}(u_y^2+q)\right)^2}=0,
\end{eqnarray*}
from which we obtain:
\begin{eqnarray}\label{alpha_1-beta_-1}
\eqalign{
\alpha_{[1]}=\frac{K(u_x^2+1)u_y}{-4(\wp(u)-\mu)u_{xx}+2\wp'(u)(u_x^2+1)}+\tilde\alpha_{[1]},\cr
\beta_{[-1]}=\frac{K(u_y^2+q)u_x}{-4(\wp(u)-\mu)u_{yy}+2\wp'(u)(u_y^2+q)}+\tilde\beta_{[-1]},}
\end{eqnarray}
where $\tilde\alpha_{[1]}$ and $\tilde\beta_{[-1]}$ are new unknown functions depending on the variables $u$, $u_x$, $u_y$. 

Thus, the dependence of the unknown functions on the variables $u_{xx}$ and $u_{yy}$ has been determined, so we can further equate the coefficients of these variables. From the remaining four equations of the system (\ref{1.13}) we obtain:
\begin{equation*}
\eqalign{
\beta_{[0]u_y}(u_y^2+q)^{3/2}-q\tilde\beta_{[-1]}\sqrt{\wp(u)-\mu}\sqrt{u_x^2+1}=0,\cr
\beta_{[1]u_y}(u_y^2+q)+\beta_{[1]}u_y=0,\cr
\alpha_{[-1]u_x}(u_x^2+1)+\alpha_{[-1]}u_x=0,\cr
\alpha_{[0]u_x}(u_x^2+1)^{3/2}-\tilde\alpha_{[1]}\sqrt{\wp(u)-\mu}\sqrt{u_y^2+q}=0.}
\end{equation*}
From which we find that the unknown functions $\beta_{[0]}$, $\beta_{[1]}$, $\alpha_{[-1]}$ and $\alpha_{[0]}$ have the form:
\begin{equation}\label{alpha_0-1-beta_01}
\eqalign{
\beta_{[0]}=\frac{\tilde\beta_{[-1]}(u,u_x)\sqrt{\wp(u)-\mu}\sqrt{u_x^2+1}u_y}{\sqrt{u_y^2+q}}
+\tilde{\beta}_{[0]}(u,u_x),\cr
\beta_{[1]}=\frac{\tilde{\beta}_{[1]}(u,u_x)}{\sqrt{u_y^2+q}},\cr
\alpha_{[-1]}=\frac{\tilde{\alpha}_{[-1]}(u,u_y)}{\sqrt{u_x^2+1}},\cr
\alpha_{[0]}=\frac{\tilde\alpha_{[1]}(u,u_y)\sqrt{\wp(u)-\mu}\sqrt{u_y^2+q}u_x}{\sqrt{u_x^2+1}}
+\tilde{\alpha}_{[0]}(u,u_y).}
\end{equation}
Here $\tilde{\alpha}_{[i]}(u,u_y)$ and $\tilde{\beta}_{[j]}(u,u_x)$ are some functions to be determined.

Let's return to the first and last equation of the system (\ref{1.13}) and rewrite these equations, taking into account (\ref{alpha_1-beta_-1}) and (\ref{alpha_0-1-beta_01}):
\begin{equation}\label{1.15}
\eqalign{
\tilde\beta_{[-1]u_y}u_{yy}+\tilde\beta_{[-1]u}u_{y}+\tilde\beta_{[-1]u_x}\sqrt{\wp(u)-\mu}\sqrt{u_x^2+1}\sqrt{u_y^2+q}\cr
\qquad -\frac{Ku_xu_y}{4(\wp(u)-\mu)}+\beta_{[1]}\alpha_{[-1]}=0,\cr
\tilde\alpha_{[1]u_x}u_{xx}+\tilde\alpha_{[1]u}u_{x}+\tilde\alpha_{[1]u_y}\sqrt{\wp(u)-\mu}\sqrt{u_x^2+1}\sqrt{u_y^2+q}\cr
\qquad -\frac{Ku_xu_y}{4(\wp(u)-\mu)}+\beta_{[1]}\alpha_{[-1]}=0.
}
\end{equation}
Since the variables $u_{xx}$ and $u_{yy}$ are independent, from the equations (\ref{1.15}) it is easy to notice the following: 
$$\tilde\beta_{[-1]}=\tilde\beta_{[-1]}(u,u_x) \quad \mbox{and} \quad \tilde\alpha_{[1]}=\tilde\alpha_{[1]}(u,u_y).$$ 
Let's continue studying the equations (\ref{1.15}). Note that subtracting one of these equations from another, we obtain the equality:
\begin{eqnarray*}
\fl \qquad \tilde\alpha_{[1]u}u_x-\tilde\beta_{[-1]u}u_y+(\tilde\alpha_{[1]u_y}-\tilde\beta_{[-1]u_x})\sqrt{\wp(u)-\mu}\sqrt{u_x^2+1}\sqrt{u_y^2+q}=0.
\end{eqnarray*}
Since the function $\tilde\beta_{[-1]u_x}$ depends only on dynamical variables $u$ and $u_y$, and the function $\tilde\alpha_{[1]u_y}$ on $u$ and $u_x$, then from the last equation, we get
\begin{equation*}
\tilde\beta_{[-1]}=\frac{\tilde{\tilde\alpha}_{[1]u}(u)\sqrt{u_x^2+1}}{\sqrt{\wp(u)-\mu}}+C_1, \qquad
\tilde\alpha_{[1]}=\frac{\tilde{\tilde\beta}_{[-1]u}(u)\sqrt{u_y^2+q}}{\sqrt{\wp(u)-\mu}}+C_2,
\end{equation*}
where $\tilde{\tilde\alpha}_{[1]u}(u)$ and $\tilde{\tilde\beta}_{[-1]u}(u)$ are functions to be found, $C_1$ and $C_2$ are some constants.
Consequently functions $\beta_{[-1]}$ and $\alpha_{[1]}$ will take the form:
\begin{equation*}
\eqalign{
 \beta_{[-1]}=\frac{K(u_y^2+q)u_x}{-4(\wp(u)-\mu)u_{yy}+2\wp'(u)(u_y^2+q)}+\frac{\tilde{\tilde\alpha}_{[1]u}(u)\sqrt{u_x^2+1}}{\sqrt{\wp(u)-\mu}}+C_1,\cr
 \alpha_{[1]}=\frac{K(u_x^2+1)u_y}{-4(\wp(u)-\mu)u_{xx}+2\wp'(u)(u_x^2+1)}+\frac{\tilde{\tilde\beta}_{[-1]u}(u)\sqrt{u_y^2+q}}{\sqrt{\wp(u)-\mu}}+C_2.}
\end{equation*}

Taking into account the above expressions, from (\ref{1.15}) we obtain the following two equations: 
\begin{equation}\label{1.15(2)}
\eqalign{
\fl \qquad \tilde\beta_{[1]}\tilde\alpha_{[-1]}=-\left(\frac{\tilde{\tilde\alpha}_{[1]uu}}{\sqrt{\wp(u)-\mu}}
-\frac{\tilde{\tilde\alpha}_{[1]u}\wp'(u)}{2(\wp(u)-\mu)^{3/2}}\right)(u_x^2+1)\sqrt{u_y^2+q}u_y\cr
-\tilde{\tilde\alpha}_{[1]u}\sqrt{u_x^2+1}(u_y^2+q)u_x
+\frac{Ku_xu_y\sqrt{u_x^2+1}\sqrt{u_y^2+q}}{4(\wp(u)-\mu)},\cr
\fl \qquad \tilde\beta_{[1]}\tilde\alpha_{[-1]}=-\left(\frac{\tilde{\tilde\beta}_{[-1]uu}}{\sqrt{\wp(u)-\mu}}
-\frac{\tilde{\tilde\beta}_{[-1]u}\wp'(u)}{2(\wp(u)-\mu)^{3/2}}\right)\sqrt{u_x^2+1}(u_y^2+q)u_x\cr
-\tilde{\tilde\beta}_{[-1]u}(u_x^2+1)\sqrt{u_y^2+q}u_y
+\frac{Ku_xu_y\sqrt{u_x^2+1}\sqrt{u_y^2+q}}{4(\wp(u)-\mu)}.
}
\end{equation}
Since the function $\tilde\beta_{[1]}$ depends only on dynamical variables $u$ and $u_x$, and the function $\tilde\alpha_{[-1]}$ on $u$ and $u_y$, then in the last two equations the coefficients for $u_y(u_x^2+1)\sqrt{u_y^2+q}$ and $u_x\sqrt{u_x^2+1}(u_y^2+q)$ they must be equal to zero, that is
$\tilde{\tilde\beta}_{[-1]u}=0$ and $\tilde{\tilde\alpha}_{[1]u}=0$. Thus, we get that $\tilde{\tilde\beta}_{[-1]}=C_3$, $\tilde{\tilde\alpha}_{[1]}=C_4$. Thus, from equations (\ref{1.15(2)}), we obtain that
\begin{equation*}
\tilde\beta_{[1]}=\frac{Ku_x\sqrt{u_x^2+1}}{4(\wp(u)-\mu)f(u)},\qquad 
\tilde\alpha_{[-1]}=f(u)u_y\sqrt{u_y^2+q},
\end{equation*}
where $f(u)$ is the function to be found. Thus, the functions $\beta_{[-1]}$ and $\alpha_{[1]}$, due to the reasoning above, take the form
\begin{equation*}
\eqalign{
\beta_{[-1]}&=\frac{K(u_y^2+q)u_x}{-4(\wp(u)-\mu)u_{yy}+2\wp'(u)(u_y^2+q)}+C_1,\cr
\alpha_{[1]}&=\frac{K(u_x^2+1)u_y}{-4(\wp(u)-\mu)u_{xx}+2\wp'(u)(u_x^2+1))}+C_2.
}
\end{equation*}
Let us substitute the found representations of the functions $\alpha_{[1]}$ and $\beta_{[-1]}$ into the third and fourth equations of the system (\ref{1.13}), we obtain
\begin{equation}\label{1.16}
\eqalign{
\fl \sqrt{u_y^2+q}\left(\frac{K}{4\sqrt{\wp(u)-\mu}f(u)}+\tilde\beta_{[0]}\right)-\frac{C_2Ku_x\sqrt{u_x^2+1}}{4(\wp(u)-\mu)f(u)}\cr
\fl \qquad +u_y\sqrt{u_x^2+1}\left(C_1\sqrt{\wp(u)-\mu}-\frac{K\wp'(u)u_x}{8(\wp(u)-\mu)^2f(u)}-\frac{Kf'(u)u_x}{4(\wp(u)-\mu)f^2(u)}\right) =0,}
\end{equation}
\begin{equation}\label{1.17}
\eqalign{
\fl u_x\sqrt{u_y^2+q}\left(f'(u)u_y+\frac{f(u)\wp'(u)u_y}{2(\wp(u)-\mu)}+C_2\sqrt{\wp(u)-\mu}\right)\cr
\fl \qquad +\sqrt{u_x^2+1}\left(qf(u)\sqrt{\wp(u)-\mu}+\tilde\alpha_{[0]}\right)+C_1f(u)u_y\sqrt{u_y^2+q}=0.}
\end{equation}
In equation (\ref{1.16}), the variable $u_y$ and the expression $\sqrt{u_y^2+q}$ can be considered independent, and in equation (\ref{1.17}) $u_x$ and $\sqrt { u_x ^ 2 +1}$ are independent, then each of these two equations splits down into the following three equations:
\begin{equation*}
\eqalign{
\frac{K}{4\sqrt{\wp(u)-\mu}f(u)}+\tilde\beta_{[0]}=0,\cr
C_1\sqrt{\wp(u)-\mu}-\frac{Ku_x\wp'(u)}{8(\wp(u)-\mu)^2f(u)}-\frac{Ku_xf'(u)}{4(\wp(u)-\mu)f^2(u)}=0,\cr
\frac{C_2Ku_x\sqrt{u_x^2+1}}{4(\wp(u)-\mu)f(u)}=0,\cr
f'(u)u_y+\frac{f(u)u_y\wp'(u)}{2(\wp(u)-\mu)}+C_2\sqrt{\wp(u)-\mu}=0,\cr
qf(u)\sqrt{\wp(u)-\mu}+\tilde\alpha_{[0]}=0,\cr
C_1f(u)u_y\sqrt{u_y^2+q}=0.}
\end{equation*}
Let us analyze the resulting equalities. Due to the fact that $f(u)\neq 0$ from the third and sixth equations we find: $C_2=0$, $C_1=0$. Then the solution to the fourth equation will be a function $f(u)$ of the form $f(u)=\frac{C_5}{\sqrt{\wp(u)-\mu}}$, where $C_5=const$. And finally the unknown functions $\tilde\beta_{[0]}$, $\tilde\alpha_{[0]}$ we find from the first and fifth equations: $\tilde\beta_{[0]}=-\frac{K}{4C_5}$, $\tilde\alpha_{[0]}=-qC_5$.

Thus all of the unknown functions are found, they are of the form (\ref{1.10}), where $\lambda:=C_5$. Theorem~1 is proved.

Now we consider some properties that follow from Theorem 1. The systems of equations (\ref{1.11}), (\ref{1.12}) can be written in the form of matrix systems
\begin{equation}\label{1.28}
\Psi_x=A\Psi, \qquad \Psi_y=B\Psi
\end{equation}
for the vector $\Psi=(v_{[1]},v_{[0]},v_{[-1]})^T$, where
\begin{eqnarray*}\label{1.29}
\fl \qquad \eqalign{
A=\left(
\begin{array}{ccc}
-b_{[0]}& h_{[0]}& 0\\
0 & -b_{[0]} & 1\\
\beta_{[1]} & \beta_{[0]} & \beta_{[-1]}-b_{[-1]}
\end{array}
\right), \qquad
B=\left(
\begin{array}{ccc}
\alpha_{[ 1]}- a_{[1]}& \alpha_{[ 0]} & \alpha_{[ -1]}\\
1 & - a_{[0]} & 0\\
0 &  h_{[-1]} & - a_{[0]} 
\end{array}
\right).}
\end{eqnarray*}
Here 
\begin{equation*}
\eqalign{
b_{[0]}=-\frac{u_y\sqrt{\wp(u)-\mu}\sqrt{u_x^2+1}}{\sqrt{u_y^2+q}}, \qquad  a_{[0]}=-\frac{u_x\sqrt{\wp(u)-\mu}\sqrt{u_y^2+q}}{\sqrt{u_x^2+1}},\cr 
h_{[0]}=-\frac{\sqrt{\wp(u)-\mu}\sqrt{u_y^2+q}u_{xx}}{(u_x^2+1)^{3/2}}+\frac{\wp'(u)\sqrt{u_y^2+q}}{2\sqrt{\wp(u)-\mu}\sqrt{u_x^2+1}},\cr 
h_{[-1]}=-\frac{q\sqrt{\wp(u)-\mu}\sqrt{u_x^2+1}u_{yy}}{(u_y^2+q)^{3/2}}+\frac{q\wp'(u)\sqrt{u_x^2+1}}{2\sqrt{\wp(u)-\mu}\sqrt{u_y^2+q}},}
\end{equation*}
\begin{equation*}
\eqalign{
\beta_{[-1]}-b_{[-1]}=\frac{u_xu_{xx}}{u_x^2+1}-\frac{u_y\sqrt{\wp(u)-\mu}\sqrt{u_x^2+1}}{\sqrt{u_y^2+q}}
+\frac{\wp'(u)u_x}{2(\wp(u)-\mu)},\cr
 \alpha_{[1]}-a_{[1]}=\frac{u_yu_{yy}}{u_y^2+q}-\frac{u_x\sqrt{\wp(u)-\mu}\sqrt{u_y^2+q}}{\sqrt{u_x^2+1}}
+\frac{\wp'(u)u_y}{2(\wp(u)-\mu)},\cr
\alpha_{[-1]}=\frac{\lambda u_y\sqrt{u_y^2+q}}{\sqrt{\wp(u)-\mu}\sqrt{u_x^2+1}}, \qquad \alpha_{[0]}=-q\lambda,\cr 
\beta_{[0]}=-\frac{K}{4\lambda}, \qquad \beta_{[1]}=\frac{Ku_x\sqrt{u_x^2+1}}{4\lambda\sqrt{\wp(u)-\mu}\sqrt{u_y^2+q}},}
\end{equation*}
where $K=\frac{g_3}{\mu}+8\mu^2$.

{\bf Corollary of Theorem 1.} 
System of equations (\ref{1.28}) is compatible if and only if function $u=u(x,y)$ is a solution to the equation (\ref{1.1(5)}). 

That is, the system of equations (\ref{1.28}) is a Lax pair for the equation (\ref{1.1(5)}). Note that this Lax pair differs from the usual one, but using suitable transformations it can be reduced to the usual Lax representation (see \S \ref{Lax}). Moreover, from (\ref{1.28}) one can derive recursion operators that describe hierarchies of higher symmetries of the equation (\ref{1.1(5)}) for both the characteristic directions of $x$ and $y$. To do this, we present the equations (\ref{1.28}) in scalar form for the function $v_{[0]}$:
\begin{equation}\label{3.5-oim}
\eqalign{
\fl \qquad \quad  v_{[0]xxx}&-\left(\frac{u_{xx}}{u_x}+\frac{2u_xu_{xx}}{u_x^2+1}\right)v_{[0]xx}\cr
&-\left(\frac{u_xu_{xxx}}{u_x^2+1}-\frac{3u_x^2u_{xx}^2}{(u_x^2+1)^2}+3\wp(u)u_x^2+(\wp(u)-\mu)-\frac{K}{4\lambda}\right)v_{[0]x}\cr
&-\left(\frac{3}{2}u_x(u_x^2+1)\wp'(u)-\frac{(\wp(u)-\mu)u_{xx}}{u_x}+\frac{Ku_{xx}}{4\lambda u_x}\right)v_{[0]}=0}
\end{equation} 
and
\begin{equation}\label{3.1-oim}
\eqalign{
\fl \qquad \quad v_{[0]yyy}&-\left(\frac{u_{yy}}{u_y}+\frac{2u_yu_{yy}}{u_y^2+q}\right)v_{[0]yy}\cr
&-\left(\frac{u_yu_{yyy}}{u_y^2+q}-\frac{3u_y^2u_{yy}^2}{(u_y^2+q)^2}+3\wp(u)u_y^2+(\wp(u)-\mu)q-\lambda q\right)v_{[0]y}\cr
&-\left(\frac{3}{2}u_y(u_y^2+q)\wp'(u)-\frac{q(\wp(u)-\mu)u_{yy}}{u_y}+\frac{q\lambda u_{yy}}{u_y}\right)v_{[0]}=0.}
\end{equation}
In operator form, the equations (\ref{3.5-oim}) and (\ref{3.1-oim}) are given as: 
\begin{eqnarray}\label{3.5}
\eqalign{\fl \left(D_x^3-\left(\frac{u_{xx}}{u_x}+\frac{2u_xu_{xx}}{u_x^2+1}\right)D_x^2-\left(\frac{u_xu_{xxx}}{u_x^2+1}-\frac{3u_x^2u_{xx}^2}{(u_x^2+1)^2}+3\wp(u)u_x^2+(\wp(u)-\mu)\right)D_x\right.\\
\fl\qquad\left.-\left(\frac{3}{2}u_x(u_x^2+1)\wp'(u)-\frac{(\wp(u)-\mu)u_{xx}}{u_x}\right)\right)v_{[0]}=-\frac{K}{4\lambda} u_xD_x\left(\frac{1}{u_x}v_{[0]}\right),}
\end{eqnarray}
\begin{eqnarray}\label{3.1}
\eqalign{\fl \left(D_y^3-\left(\frac{u_{yy}}{u_y}+\frac{2u_yu_{yy}}{u_y^2+q}\right)D_y^2-\left(\frac{u_yu_{yyy}}{u_y^2+q}-\frac{3u_y^2u_{yy}^2}{(u_y^2+q)^2}+3\wp(u)u_y^2+(\wp(u)-\mu)q\right)D_y\right.\\
\fl \qquad \left.-\left(\frac{3}{2}u_y(u_y^2+q)\wp'(u)-\frac{(\wp(u)-\mu)u_{yy}}{u_y}\right)\right)v_{[0]}=-q\lambda u_yD_y\left(\frac{1}{u_y}v_{[0]}\right).}
\end{eqnarray}
Let us transform the last two equalities as follows. We will multiply the equation (\ref{3.5}) by $u_xD_x^{-1}\frac{1}{u_x}$ and (\ref{3.1}) by $u_yD_y^{-1}\frac{1}{u_y }$, respectively, then we get 
\begin{equation*}
R_xv_{[0]}=-\frac{K}{4\lambda} v_{[0]} \quad \mbox{and} \quad R_yv_{[0]}=-q\lambda v_{[0]},
\end{equation*}
where $R_x$ and $R_y$ are operators of the form:
\begin{eqnarray}\label{R-1}
\eqalign{
\fl \qquad R_x=D_x^2-\frac{2u_xu_{xx}}{u_x^2+1}D_x-\left(\frac{u_x^2u_{xx}^2}{(u_x^2+1)^2}+3\wp(u)u_x^2+(\wp(u)-\mu)\right)\\
\fl \qquad \qquad +u_xD_x^{-1}\left(\frac{u_{xx}u_{xxx}}{u_x(u_x^2+1)}+\frac{u_{xx}\left(2u_xu_{xxx}+u_{xx}^2\right)}{(u_x^2+1)^2}-\frac{4u_x^2u_{xx}^3}{(u_x^2+1)^3}\right.\\
\fl \qquad \qquad \left.+\frac{1}{2}\left(3u_x^2-1\right)\wp'(u)+3\wp(u)u_{xx}\right),\\
\fl \qquad R_y=D_y^2-\frac{2u_yu_{yy}}{u_y^2+q}D_y-\left(\frac{u_y^2u_{yy}^2}{(u_y^2+q)^2}+3\wp(u)u_y^2+(\wp(u)-\mu)q\right)\\
\fl \qquad \qquad +u_yD_y^{-1}\left(\frac{u_{yy}u_{yyy}}{u_y(u_y^2+q)}+\frac{u_{yy}\left(2u_yu_{yyy}+u_{yy}^2\right)}{(u_y^2+q)^2}-\frac{4u_y^2u_{yy}^3}{(u_y^2+q)^3}\right.\\
\fl \qquad \qquad \left.+\frac{1}{2}\left(3u_y^2-q\right)\wp'(u)+3\wp(u)u_{yy}\right).}
\end{eqnarray}

{\bf Proposition 1.} 
Operators $R_x$ and $R_y$ given in (\ref{R-1}) define recursion operators generating hierarchies of the symmetries in the directions of $x$ and respectively $y$ for equation (\ref{1.1(5)}).

Indeed, if we apply the operators $R_x$ and $R_y$ to the right-hand sides of the classical symmetries $u_{t_1}=u_x$ and respectively $u_{\tau_1}=u_y$ of equation (\ref{1.1(5)}), then we get:
\begin{eqnarray*}
& R_x(u_x)=u_{xxx}-\frac{3u_xu^2_{xx}}{2(u_x^2+1)}-\frac{3}{2}\wp(u)u_x(u_x^2+1),\\
& R_y(u_y)=u_{yyy}-\frac{3u_yu^2_{yy}}{2(u_y^2+q)}-\frac{3}{2}\wp(u)u_y(u_y^2+q),
\end{eqnarray*}
which are the first members of the hierarchies of higher symmetries of equation (\ref{1.1(5)}):
\begin{eqnarray}
u_t&=u_{xxx}-\frac{3u_xu^2_{xx}}{2(u_x^2+1)}-\frac{3}{2}\wp(u)u_x(u_x^2+1), \label{symm1} \\ 
u_\tau&=u_{yyy}-\frac{3u_yu^2_{yy}}{2(u_y^2+q)}-\frac{3}{2}\wp(u)u_y(u_y^2+q). \label{symm2}
\end{eqnarray}
Equations (\ref{symm1}) and (\ref{symm2}) were found in the work \cite{SokMesh}, where it was also proven that they themselves are integrable equations.

{\bf Proof of the Proposition 1.}
To prove Proposition 1, we use the fact that the operator $R$ is a recursion operator of a hyperbolic equation of the form (\ref{Iy.1}), if condition 
\begin{equation}\label{condR}
\frac{d}{dt}R=\left[F_*,R\right],
\end{equation}
is satisfied, where $F_ *$ is the linearization operator of some higher symmetry (see \cite{Olver, IbragimovShabat, Sokolov}). As $F_*$ we can take the linearization operator of equation (\ref{symm1}), since this is the simplest higher symmetry (\ref{1.1(5)}) in the direction of $x$:
\begin{eqnarray*}
\fl \quad F_*=D_x^3-\frac{3u_xu_{xx}}{u_x^2+1}D_x^2+\left(\frac{3u_{xx}^2(u_x^2-1)}{2(u_x^2+1)^2}-\frac{3}{2}\wp(u)(3u_x^2+1)\right)D_x-\frac{3}{2}\wp'(u)u_x(u_x^2+1).
\end{eqnarray*}
Let us substitute the operator $R_x$ from (\ref{R-1}) and the operator $F_*$ into formula (\ref{condR}): 
\begin{eqnarray*}
\fl \quad
\frac{d}{dt}\left[D_x^2-\frac{2u_xu_{xx}}{u_x^2+1}D_x-\left(\frac{u_x^2u_{xx}^2}{(u_x^2+1)^2}+3\wp(u)u_x^2+(\wp(u)-\mu)\right)\right.\\
\fl \quad
\left.+u_xD_x^{-1}\left(\frac{u_{xx}u_{xxx}}{u_x(u_x^2+1)}+\frac{u_{xx}\left(2u_xu_{xxx}+u_{xx}^2\right)}{(u_x^2+1)^2}-\frac{4u_x^2u_{xx}^3}{(u_x^2+1)^3}
+\frac{1}{2}\left(3u_x^2-1\right)\wp'(u)+3\wp(u)u_{xx}\right)\right]\\
\fl \quad
 =\left[D_x^3-\frac{3u_xu_{xx}}{u_x^2+1}D_x^2+\left(\frac{3u_{xx}^2(u_x^2-1)}{2(u_x^2+1)^2}-\frac{3}{2}\wp(u)(3u_x^2+1)\right)D_x-\frac{3}{2}\wp'(u)u_x(u_x^2+1),\right.\\
\fl \quad
\left.D_x^2-\frac{2u_xu_{xx}}{u_x^2+1}D_x-\left(\frac{u_x^2u_{xx}^2}{(u_x^2+1)^2}+3\wp(u)u_x^2+(\wp(u)-\mu)\right)\right.\\
\fl \quad
\left.+u_xD_x^{-1}\left(\frac{u_{xx}u_{xxx}}{u_x(u_x^2+1)}+\frac{u_{xx}\left(2u_xu_{xxx}+u_{xx}^2\right)}{(u_x^2+1)^2}-\frac{4u_x^2u_{xx}^3}{(u_x^2+1)^3}
+\frac{1}{2}\left(3u_x^2-1\right)\wp'(u)+3\wp(u)u_{xx}\right)\right].
\end{eqnarray*}
The validity of equality is verified by direct calculation. Due to the cumbersome calculations, we do not provide them. The same way, it is verified that $R_y$ is a recursion operator of equation (\ref{1.1(5)}) in the direction of $y$.

\section{Construction of the Lax pair of the usual form}\label{Lax}

Let us construct the Lax pair for equation (\ref{1.1(5)}). To do this, we need to use four equations: the equation (\ref{1.1(5)}), the linearization of the equation (\ref{1.1(5)}) and the two third-order equations that we built in the previous section (\ref {3.5-oim}) and (\ref{3.1-oim}). Let's start with the equation (\ref {3.5-oim}), which for convenience we rewrite in the form
\begin{equation}\label{vxxx}
 v_{xxx}=Av_{xx}+Bv_{x}+Cv,
\end{equation} 
where $A=\frac{u_{xx}}{u_x}+\frac{2u_xu_{xx}}{u_x^2+1}$, $B=\frac{u_xu_{xxx}}{u_x^2+1}-\frac{3u_x^2u_{xx}^2}{(u_x^2+1)^2}+3\wp(u)u_x^2+(\wp(u)-\mu)-\frac{K}{4\lambda}$, \linebreak $C=\frac{3}{2}\wp'(u)u_x(u_x^2+1)-\frac{(\wp(u)-\mu)u_{xx}}{u_x}+\frac{K u_{xx}}{4\lambda u_x}$.
First we lower the order of this equation. In other words, we will look for a connection condition of the form 
\begin{equation}\label{vxx}
v_{xx}=F(u,u_x,u_{xx},v,v_x)
\end{equation}
that satisfies equation
\begin{equation*}
\left.Av_{xx}+Bv_{x}+Cv-D_x(F(u,u_x,u_{xx},v,v_x))\right|_{(\ref{vxxx}),(\ref{vxx})}=0,
\end{equation*}
or in expanded form
\begin{equation}\label{satis-cond}
\eqalign{
\fl \qquad \frac{u_{xx}(3u_x^2+1)}{u_x(u_x^2+1)}F+\frac{u_xu_{xxx}}{u_x^2+1}v_{x}-\frac{3u_x^2u_{xx}^2}{(u_x^2+1)^2}v_{x}+3\wp(u)u_x^2v_{x}+(\wp(u)-\mu)v_{x}\cr
\fl \qquad \qquad -\frac{K}{4\lambda}v_{x}+\frac{(K-4\lambda(\wp(u)-\mu))u_{xx}}{4\lambda u_x}v+\frac{3}{2}\wp'(u)u_x(u_x^2+1)v\cr
\fl \qquad \qquad -F_uu_x-F_{u_x}u_{xx}-F_{u_{xx}}u_{xx}-F_vv_x-F_{v_x}F=0.}
\end{equation}
Equation (\ref{satis-cond}) is the equation for finding the unknown function $F=F(u,u_x,u_{xx},v,v_x)$. Note that the variables $u, u_x, u_{xx}, \ldots$ are independent, so the problem of finding the function $F$ is reduced to solving an overdetermined system of equations. First, in the equation (\ref{satis-cond}), we equate the coefficients of the highest derivative $u_{xxx}$:
\begin{equation*}
F_{u_{xx}}-\frac{u_xv_x}{u^2_x+1}=0.
\end{equation*}
From the last equation it follows that the function $F$ depends linearly on the variable~$u_{xx}$:
\begin{equation}\label{F-F1}
F(u,u_x,u_{xx},v,v_x)=\frac{u_xv_x}{u^2_x+1}u_{xx}+F_1(u,u_x,v,v_x),
\end{equation}
where $F_1=F_1(u,u_x,v,v_x)$ is the new sought function.
Let's substitute the found value (\ref{F-F1}) into the equation (\ref{satis-cond}) and get:
\begin{equation}\label{satis-cond-2}
\eqalign{
\fl \qquad \left((F_1)_{u_x}+\frac{u_xv_x}{u^2_x+1}(F_1)_{v_x}-\frac{2u_x^2+1}{u_x(u_x^2+1)}F_1+\frac{(\wp(u)-\mu)v}{u_x}-\frac{Kv}{4u_x\lambda}\right)u_{xx}\cr
\fl \qquad \qquad + F_1(F_1)_{v_x}+(F_1)_{u}u_x+(F_1)_{v}v_x-\frac{3}{2}u_x(u_x^2+1)v\wp'(u)\cr
\fl \qquad \qquad - (3u_x^2+1)\wp(u)v_x+\mu v_x+\frac{Kv_x}{4\lambda}=0. }
\end{equation}
Since the dependence of the function $F$ on $u_{xx}$ has been determined, we equate the coefficients of this variable in the last equality and find that the function $F_1$ has the form: 
\begin{equation}\label{F1-F2}
\eqalign{
F_1(u,u_x,v,v_x)=&u_x\sqrt{u_x^2+1}F_2\left(u,v,\frac{v_x}{\sqrt{u_x^2+1}}\right)\cr
&+(u_x^2+1)\left(\wp(u)-\mu-\frac{K}{4\lambda}\right)v,}
\end{equation}
where $F_2=F_2\left(u,v,\frac{v_x}{\sqrt{u_x^2+1}}\right)$ is the new sought function, satisfying the following equation
\begin{equation}\label{satis-cond-3}
\eqalign{
\fl \qquad\quad F_2(F_2)_h+(F_2)_u+\frac{v\sqrt{u_x^2+1}}{u_x}\left(\wp(u)-\mu-\frac{K}{4\lambda}\right)(F_2)_h\cr
\fl \qquad\qquad +\frac{h\sqrt{u_x^2+1}}{u_x}(F_2)_v-\frac{v\wp'(u)\sqrt{u_x^2+1}}{2u_x}-h\left(4\wp(u)+\mu+\frac{K}{4\lambda}\right)=0.
}
\end{equation}
Here, for convenience, notation $h:=\frac{v_x}{\sqrt{u_x^2+1}}$ is introduced, i.e. we excluded variable $v_x$ due to $v_x=h\sqrt{u_x^2+1}$. Note that equation (\ref{satis-cond-3}) is obtained from (\ref{satis-cond-2}) due to the found equality (\ref{F1-F2}). Let's focus on the equation (\ref{satis-cond-3}). Now our unknown function $F_2$ depends on the variables $u$, $v$ and $h$, i.e. the variable $u_x$ is independent and we can equate the coefficients of this variable. Namely, let’s collect the coefficients of the factor $\frac{\sqrt{u_x^2+1}}{u_x}$: 
\begin{equation*}
 h(F_2)_v+v\left(\wp(u)-\mu-\frac{K}{4\lambda}\right)(F_2)_h-\frac{1}{2}v\wp'(u)=0
\end{equation*}
from which we obtain
\begin{equation*}
\fl \qquad\qquad F_2(u,v,h)=\frac{2h\lambda \wp'(u)}{4\lambda\wp(u)-4\lambda\mu-K}+F_3\left(u,h^2-v^2(\wp(u)-\mu)+\frac{v^2K}{4\lambda}\right)
\end{equation*}
or, returning to the old variables $h=\frac{v_x}{\sqrt{u_x^2+1}}$, we have
\begin{equation*}
\eqalign{
F_2\left(u,v,\frac{v_x}{\sqrt{u_x^2+1}}\right)=&\frac{2\lambda \wp'(u)v_x}{\sqrt{u_x^2+1}\left(4\lambda\wp(u)-4\lambda\mu-K\right)}\cr
&+F_3\left(u,\frac{v^2_x}{u_x^2+1}-v^2(\wp(u)-\mu)+\frac{v^2K}{4\lambda}\right).}
\end{equation*}
Next, we substitute the found value of the function $F_2$ into the equation (\ref{satis-cond-3}) and make a change of variables $\theta=\frac{v^2_x}{u_x^2+1}-v^2(\wp(u)-\mu)+\frac{v^2K}{4\lambda}$, namely, we exclude the variable $v_x$ in the entire equation due to this change. Then we obtain the equation for the unknown function $F_3(u,\theta)$: 
\begin{equation}\label{satis-cond-4}
\eqalign{
\left(\frac{1}{\sqrt{\lambda}}F_3(F_3)_\theta-\frac{K\left(48\lambda^2\mu^2+12K\lambda\mu+K^2-4g_2\lambda^2\right)}{8\lambda^{\frac{3}{2}}\left(-4\lambda\wp(u)+4\lambda\mu+K\right)^2}\right)S\cr
\qquad \qquad +(F_3)_u-\frac{2\lambda\wp'(u)\left(F_3+2\theta (F_3)_\theta\right)}{-4\lambda\wp(u)+4\lambda\mu+K}=0,
}
\end{equation}
where $S=\pm\sqrt{4\lambda\wp(u)v^2-4\lambda\mu v^2-Kv^2+4\lambda\theta}$. Since the expression under the root sign contains the independent variable $v$, the equation (\ref{satis-cond-4}) can be divided into the following two equalities: 
\begin{equation}\label{satis-cond-4-2}
\eqalign{
1) \quad F_3(F_3)_\theta-\frac{K\left(48\lambda^2\mu^2+12K\lambda\mu+K^2-4g_2\lambda^2\right)}{8\lambda\left(-4\lambda\wp(u)+4\lambda\mu+K\right)^2}=0,\cr
2) \quad (F_3)_u-\frac{2\lambda\wp'(u)\left(F_3+2\theta (F_3)_\theta\right)}{-4\lambda\wp(u)+4\lambda\mu+K}=0.}
\end{equation}
From the second equation of the system (\ref{satis-cond-4-2}) we find:
\begin{equation}\label{F3-F4}
F_3(u,\theta)=\frac{1}{\sqrt{-4\lambda\wp(u)+4\lambda\mu+K}} \, F_4\left(\frac{\theta}{-4\lambda\wp(u)+4\lambda\mu+K}\right).
\end{equation}
Due to the found value (\ref{F3-F4}), the first equation of the system (\ref{satis-cond-4-2}) is written in the form
\begin{equation}\label{satis-cond-4-first}
48\lambda^2\mu^2 K+12\lambda\mu K^2-4\lambda^2 g_2K+K^3-8F_4(F_4)_\eta=0,
\end{equation}
where $\eta=\frac{\theta}{-4\lambda\wp(u)+4\lambda\mu+K}$. Solving the equation (\ref{satis-cond-4-first}) and returning to the variables $u$, $u_x$, $v$ and $v_x$ we find that the function $F_4$ has the form:
\begin{equation*}
\eqalign{
F_4\left(\frac{v^2_x}{\left(u_x^2+1\right)\left(K-4\lambda(\wp(u)-\mu)\right)}+\frac{v^2}{4\lambda}\right)\cr
\qquad =\frac{\varepsilon \sqrt{(u_x^2+1)\left(K-4\lambda(\wp(u)-\mu)\right)\left(K_1v^2-\lambda^3C\right)+4\lambda K_1v_x^2}}{4\lambda\sqrt{u_x^2+1}\sqrt{K-4\lambda(\wp(u)-\mu)}},
}
\end{equation*}
where $K_1=K^2\left(4\lambda^2+12\mu\lambda+K\right)$, $\varepsilon=\pm 1$.

Finally, collecting all the calculations, we find that our desired function $F(u,u_x,u_{xx},v,v_x)$ has the form
\begin{equation*}
\eqalign{
\fl F(u,u_x,u_{xx},v,v_x)=\left(\frac{u_xu_{xx}}{u_x^2+1}+\frac{2\lambda u_x\wp'(u)}{4\lambda(\wp(u)-\mu)-K}\right)v_x\cr
\qquad +\frac{(u_x^2+1)\left(4\lambda(\wp(u)-\mu)-K\right)}{4\lambda}v\cr
\qquad +\frac{\varepsilon u_x\sqrt{(u_x^2+1)\left(K-4\lambda(\wp(u)-\mu)\right)\left(K_1v^2-\lambda^3C\right)+4\lambda K_1v_x^2}}{4\lambda(K-4\lambda(\wp(u)-\mu))}.
}
\end{equation*}
Thus, the connection condition (\ref{vxx}) has the following form:
\begin{equation}\label{vxx-final}
\eqalign{
\fl v_{xx}=\left(\frac{u_xu_{xx}}{u_x^2+1}+\frac{2\lambda u_x\wp'(u)}{4\lambda(\wp(u)-\mu)-K}\right)v_x\cr
\qquad +\frac{(u_x^2+1)\left(4\lambda(\wp(u)-\mu)-K\right)}{4\lambda}v\cr
\qquad +\frac{\varepsilon u_x\sqrt{(u_x^2+1)\left(K-4\lambda(\wp(u)-\mu)\right)\left(K_1v^2-\lambda^3C\right)+4\lambda K_1v_x^2}}{4\lambda(K-4\lambda(\wp(u)-\mu))}.
}
\end{equation}
In order to construct a Lax pair of the equation (\ref{1.1(5)}), we need to derive one more equality connecting the variables $v_y$, $v_x$, $v$, as well as the variable $u$ and its derivatives with respect to $ x$ and $y$. To do this, we differentiate the equation (\ref{vxx-final}) with respect to $y$ and replace the mixed variables $v_{xy}$ and $v_{xxy}$ by virtue of the linearized equation (\ref{lin5}), and the variables  $u_{xx}$ and $u_{xxy}$ by virtue of the original equation (\ref{1.1(5)}). Then, after transforming the resulting equality, we find:
\begin{equation}\label{vy-final}
\eqalign{
\fl v_y=-\frac{2\lambda \wp'(u)\sqrt{u_y^2+q}}{\sqrt{\wp(u)-\mu}\sqrt{u_x^2+1}\left(K-4\lambda(\wp(u)-\mu)\right)}v_x\cr
\fl \quad -\frac{\varepsilon\sqrt{u_y^2+q}\sqrt{\wp(u)-\mu}\sqrt{(u_x^2+1)\left(K-4\lambda(\wp(u)-\mu)\right)\left(K_1v^2-\lambda^3C\right)+4\lambda K_1v_x^2}}{K\sqrt{u_x^2+1}\left(K-4\lambda(\wp(u)-\mu)\right)}.}
\end{equation}
Let's rewrite linearization (\ref{lin5}) using equation (\ref{vy-final}):
\begin{equation}\label{vxy-final}
\eqalign{
\fl v_{xy}=\left(\frac{u_x\sqrt{u_y^2+q}\sqrt{\wp(u)-\mu}}{\sqrt{u_x^2+1}}-\frac{2\lambda u_y\wp'(u)}{K-4\lambda(\wp(u)-\mu)}\right)v_x\cr
\fl \qquad +\frac{\sqrt{u_x^2+1}\sqrt{u_y^2+q}\wp'(u)}{2\sqrt{\wp(u)-\mu}}v\cr
\fl \qquad -\frac{\varepsilon u_y\sqrt{\wp(u)-\mu}\sqrt{(u_x^2+1)\left(K-4\lambda(\wp(u)-\mu)\right)\left(K_1v^2-\lambda^3C\right)+4\lambda K_1v_x^2}}{K\left(K-4\lambda(\wp(u)-\mu)\right)}.}
\end{equation}

Note that the triple of equations (\ref{vxx-final}), (\ref{vy-final}), (\ref{vxy-final}) is consistent on solutions of the given equation (\ref{1.1(5)}).

Now, using this joint triple of equations (\ref{vxx-final}), (\ref{vy-final}), (\ref{vxy-final}), we will find the usual Lax pair for the equation (\ref{1.1(5)}). Let's set $C=0$ and make the following change of variables in these equations:
\begin{eqnarray}
v=4\sqrt{\lambda} \varphi(x,y)\psi(x,y), \label{ch-1} \\
v_x=\sqrt{u_x^2+1}\sqrt{K-4\lambda(\wp(u)-\mu)}\left((\varphi(x,y))^2-(\psi(x,y))^2\right). \label{ch-2}
\end{eqnarray}
First, we differentiate (\ref{ch-1}) with respect to $x$ and subtract (\ref{ch-2}) from it:
\begin{eqnarray}\label{eq1}
\fl \qquad \quad 4\sqrt{\lambda} \left(\varphi_x\psi+\varphi\psi_x\right)-\sqrt{u_x^2+1}\sqrt{K-4\lambda(\wp(u)-\mu)}\left(\varphi^2-\psi^2\right)=0. 
\end{eqnarray}
Next, we will derive another equation connecting the variables $\varphi_x$ and $\psi_x$. To do this, let us focus on equalities (\ref{vxx-final}) and (\ref{ch-2}). We differentiate (\ref{ch-2}) with respect to $x$ and subtract from it equation (\ref{vxx-final}) in which we replace the variables $v$ and $v_x$ by virtue of equalities (\ref{ch-1}) and (\ref{ch-2}). After simple transformations we get:
\begin{eqnarray}\label{eq2}
\eqalign{
\varphi\varphi_x-\psi\psi_x&+\frac{\sqrt{u_x^2+1}\sqrt{K-4\lambda(\wp(u)-\mu)}}{2\sqrt{\lambda}}\varphi\psi\cr
&-\frac{\varepsilon u_x K\sqrt{4\lambda^2+12\mu\lambda+K}\left(\varphi^2+\psi^2\right)}{4\sqrt{\lambda}(K-4\lambda(\wp(u)-\mu))}=0.}
\end{eqnarray}
Now, combining equations (\ref{eq1}) and (\ref{eq2}), we obtain the following two equations:
\begin{eqnarray}\label{px-1}
\eqalign{
\fl \qquad \varphi_x=-\frac{\varepsilon u_xK\sqrt{4\lambda^2+12\mu\lambda+K}}{4\sqrt{\lambda}(K-4\lambda(\wp(u)-\mu))}\varphi-\frac{\sqrt{u_x^2+1}\sqrt{K-4\lambda(\wp(u)-\mu)}}{4\sqrt{\lambda}}\psi,  \\
\fl \qquad \psi_x=\frac{\sqrt{u_x^2+1}\sqrt{K-4\lambda(\wp(u)-\mu)}}{4\sqrt{\lambda}}\varphi+\frac{\varepsilon u_xK\sqrt{4\lambda^2+12\mu\lambda+K}}{4\sqrt{\lambda}(K-4\lambda(\wp(u)-\mu))}\psi. }
\end{eqnarray}
In a similar way, we derive equations for the variables $\varphi_y$ and $\psi_y$. Here we will use substitutions (\ref{ch-1}), (\ref{ch-2}) and equations (\ref{vy-final}), (\ref{vxy-final}). To obtain the first equation, we differentiate (\ref{ch-1}) with respect to $y$ and subtract equation (\ref{vy-final}) from it, and in order to find the second equality, we differentiate (\ref{ch-2}) with respect to $y$ and subtract equation (\ref{vxy-final}) from it. As a result, we obtain the following two equations: 
\begin{eqnarray*}
\eqalign{
\varphi_y\psi+\varphi\psi_y&+\frac{\wp'(u)\sqrt{\lambda}\sqrt{u_y^2+q}}{2\sqrt{\wp(u)-\mu}\sqrt{K-4\lambda(\wp(u)-\mu)}}\left(\varphi^2-\psi^2\right)\cr
&+\frac{\varepsilon \sqrt{4\lambda^2+12\mu\lambda+K}\sqrt{u_y^2+q}\sqrt{\wp(u)-\mu}\left(\varphi^2+\psi^2\right)}{2\sqrt{K-4\lambda(\wp(u)-\mu)}}=0,}
\end{eqnarray*}
\begin{eqnarray*}
\eqalign{
\varphi\varphi_y+\psi\psi_y
&-\frac{\wp'(u)\sqrt{\lambda}\sqrt{u_y^2+q}}{\sqrt{\wp(u)-\mu}\sqrt{K-4\lambda(\wp(u)-\mu)}}\varphi\psi\cr
&+\frac{\varepsilon u_y\sqrt{\lambda}\sqrt{4\lambda^2+12\mu\lambda+K}\sqrt{\wp(u)-\mu}\left(\varphi^2+\psi^2\right)}{\left(K-4\lambda(\wp(u)-\mu)\right)}=0,}
\end{eqnarray*}
from which the equations for $\varphi_y$ and $\psi_y$ are easily derived:
\begin{eqnarray}\label{py-1}
\eqalign{
\fl \qquad \varphi_y=-\frac{\varepsilon u_y(\wp(u)-\mu)\sqrt{\lambda}\sqrt{4\lambda^2+12\mu\lambda+K}}{K-4\lambda(\wp(u)-\mu)}\varphi\cr
\fl \qquad \qquad +\frac{\sqrt{u_y^2+q}\sqrt{\wp(u)-\mu}}{2\sqrt{K-4\lambda(\wp(u)-\mu)}}\left(\frac{\wp'(u)\sqrt{\lambda}}{\wp(u)-\mu}-\varepsilon\sqrt{4\lambda^2+12\mu\lambda+K}\right)\psi,  \\
\fl \qquad \psi_y=-\frac{\sqrt{u_y^2+q}\sqrt{\wp(u)-\mu}}{2\sqrt{K-4\lambda(\wp(u)-\mu)}}\left(\frac{\wp'(u)\sqrt{\lambda}}{\wp(u)-\mu}+\varepsilon\sqrt{4\lambda^2+12\mu\lambda+K}\right)\varphi\cr
\fl \qquad \qquad +\frac{\varepsilon u_y(\wp(u)-\mu)\sqrt{\lambda}\sqrt{4\lambda^2+12\mu\lambda+K}}{K-4\lambda(\wp(u)-\mu)}\psi. }
\end{eqnarray}

{\bf Theorem 2.} System of equations (\ref{px-1}), (\ref{py-1}) is consistent only and only if equation (\ref{1.1(5)}) is satisfied, i.e. it defines a Lax pair for equation (\ref{1.1(5)}).

{\bf Proof.} The proof of Theorem 2 is obtained from the following compatibility conditions by direct calculation:
\begin{eqnarray*}
\left.D_y(\varphi_x)-D_x(\varphi_y)\right|_{(\ref{px-1}), (\ref{py-1})}=0,\\
\left.D_y(\psi_x)-D_x(\psi_y)\right|_{(\ref{px-1}), (\ref{py-1})}=0.
\end{eqnarray*}

Let us simplify the Lax pair (\ref{px-1}), (\ref{py-1}). To do this, we parametrize the spectral parameter $\lambda$ as follows: $\lambda=p^2sn^2(z;k)$. First we rewrite the radical expression $4\lambda^2+12\mu\lambda+K$ in the Lax pair (\ref{px-1}), (\ref{py-1}) taking into account the replacement:
\begin{eqnarray*}
\fl \qquad \left.4\lambda^2+12\mu\lambda+K\right|_{\lambda=p^2sn^2(z;k)}=K\left(\frac{4}{K}p^4sn^4(z;k)+\frac{12\mu}{K}p^2sn^2(z;k)+1\right).
\end{eqnarray*}
Let us put $\frac{4}{K}p^4=k^2$, $\frac{12\mu}{K}p^2=-(k^2+1)$ in the last expression, where $p$ is the solution to the equation $\frac{4}{K}p^4+\frac{12\mu}{K}p^2+1=0$. Then we get:
\begin{equation*}
\fl \qquad \eqalign{
K\left(k^2sn^4(z;k)-(k^2+1)sn^2(z;k)+1\right)=K\left(1-sn^2(z;k)\right)\left(1-k^2sn^2(z;k)\right)\cr
\qquad \qquad =K\left(\frac{d sn(z;k)}{d z}\right)^2=K\, cn^2(z;k)\,dn^2(z;k).}
\end{equation*}
Now we can rewrite the Lax pair taking into account the latest calculations:
\begin{eqnarray*} 
\fl\quad\left\{\begin{array}{l}
\displaystyle{\varphi_x=-\frac{u_xK^{3/2}\,sn'(z;k))}{4\, p\, sn(z;k)(K-4\, p^2\,sn^2(z;k)(\wp(u)-\mu))}\varphi}\cr
\qquad \qquad \displaystyle{-\frac{\sqrt{u_x^2+1}\sqrt{K-4\, p^2\,sn^2(z;k)(\wp(u)-\mu)}}{4\,p\, sn(z;k)}\psi, }\\ \\
\displaystyle{\psi_x=\frac{\sqrt{u_x^2+1}\sqrt{K-4\, p^2\,sn^2(z;k)(\wp(u)-\mu)}}{4\,p\, sn(z;k)}\varphi}\cr
\qquad \qquad \displaystyle{+\frac{u_xK^{3/2}\,sn'(z;k)}{4\,p\, sn(z;k)(K-4\, p^2\,sn^2(z;k)(\wp(u)-\mu))}\psi,} 
\end{array}\right.\\ \\
\fl\quad\left\{\begin{array}{l}
\displaystyle{\varphi_y=-\frac{ u_y(\wp(u)-\mu)\,sn(z;k)\,sn'(z;k)\sqrt{K}\,p}{(K-4\, p^2\,sn^2(z;k)(\wp(u)-\mu))}\varphi}\cr
\qquad \qquad \displaystyle{+\frac{\sqrt{u_y^2+q}\sqrt{\wp(u)-\mu}}{2\sqrt{K-4\,p^2\,sn^2(z;k)(\wp(u)-\mu)}}\left(\frac{p\, sn(z;k)\wp'(u)}{\wp(u)-\mu}-\sqrt{K}\,sn'(z;k)\right)\psi,}\\ \\
\displaystyle{\psi_y=-\frac{\sqrt{u_y^2+q}\sqrt{\wp(u)-\mu}}{2\sqrt{K-4\, p^2\,sn^2(z;k)(\wp(u)-\mu)}}\left(\frac{p\, sn(z;k)\wp'(u)}{\wp(u)-\mu}+\sqrt{K}\,sn'(z;k)\right)\varphi}\cr
\qquad \qquad \displaystyle{+\frac{ u_y(\wp(u)-\mu)\,sn(z;k)\,sn'(z;k)\sqrt{K}\,p}{(K-4\, p^2\,sn^2(z;k)(\wp(u)-\mu))}\psi.}
\end{array}\right.
\end{eqnarray*}

\section{General form of equation (\ref{1.1(5)})}

Equation (\ref{1.1(5)}) in the general case can be written through the Jacobi function $sn(u;k)$: 
\begin{eqnarray}
u_{xy}=\frac{1}{sn(u;k)}\sqrt{u_x^2+1}\sqrt{u_y^2+q}. \label{gen-case}
\end{eqnarray}
This equation first appeared in the work of A.B. Borisov and S.A. Zykov \cite{Borisov}. Below, we present the Lax pair and recursion operators for both characteristic directions for equation (\ref{gen-case}). To our knowledge, these objects have not been found previously. 

The recursion operators for equation (\ref{gen-case}) have the form: 
\begin{eqnarray}\label{R-2}
\eqalign{
\fl \quad
R_x=D_x^2-\frac{2u_xu_{xx}}{u_x^2+1}D_x+\left(-\frac{u_x^2u_{xx}^2}{(u_x^2+1)^2}+\frac{(k^2+1)sn^2(u;k)-3}{sn^2(u;k)}u_x^2-\frac{1}{sn^2(u;k)}\right)\\
\fl \qquad \quad +u_xD_x^{-1}\left(\frac{u_{xx}u_{xxx}}{u_x(u_x^2+1)}+\frac{u_{xx}\left(2u_xu_{xxx}+u_{xx}^2\right)}{(u_x^2+1)^2}-\frac{4u_x^2u_{xx}^3}{(u_x^2+1)^3}\right.\\
\fl \qquad \quad \left.-\frac{(k^2+1)sn^2(u;k)-3}{sn^2(u;k)}u_{xx}+\frac{cn(u;k)dn(u;k)(-3u_x^2+1)}{sn^3(u;k)}\right),\\
\fl \quad
R_y=D_y^2-\frac{2u_yu_{yy}}{u_y^2+q}D_y+\left(-\frac{u_y^2u_{yy}^2}{(u_y^2+q)^2}+\frac{(k^2+1)sn^2(u;k)-3}{sn^2(u;k)}u_y^2-\frac{q}{sn^2(u;k)}\right)\\
\fl \qquad \quad +u_yD_y^{-1}\left(\frac{u_{yy}u_{yyy}}{u_y(u_y^2+q)}+\frac{u_{yy}\left(2u_yu_{yyy}+u_{yy}^2\right)}{(u_y^2+q)^2}-\frac{4u_y^2u_{yy}^3}{(u_y^2+q)^3}\right.\\
\fl \qquad \quad \left.-\frac{(k^2+1)sn^2(u;k)-3}{sn^2(u;k)}u_{yy}+\frac{cn(u;k)dn(u;k)(-3u_y^2+q)}{sn^3(u;k)}\right).}
\end{eqnarray}

{\bf Proposition 2.} Operators $R_x$ and $R_y$ given in (\ref{R-2}) define recursion operators generating hierarchies of the symmetries in the directions of $x$ and respectively $y$ for equation (\ref{gen-case}).

It can be shown that by applying these operators to the generators of the classical symmetries $u_{t_1}=u_x$ and $u_{\tau_1}=u_y$ of equation (\ref{gen-case}) we obtain the first members of the hierarchies of higher symmetries of equation (\ref{gen-case}):
\begin{eqnarray*}
u_t&=u_{xxx}-\frac{3u_xu^2_{xx}}{2(u_x^2+1)}-\frac{1}{2}\left(\frac{3}{sn(u;k)^2}-(k^2+1)\right)u_x(u_x^2+1), \\ 
u_\tau&=u_{yyy}-\frac{3u_yu^2_{yy}}{2(u_y^2+q)}-\frac{1}{2}\left(\frac{3}{sn(u;k)^2}-(k^2+1)\right)u_y(u_y^2+q).
\end{eqnarray*}
Previously, in the work \cite{SokMesh}, it was proven that these equations are integrable.

The Lax pair for equation (\ref{gen-case}) has the form:
\begin{eqnarray*} 
\fl\quad\left\{\begin{array}{l}\displaystyle{\varphi_x=-\frac{u_x}{2}\left(\frac{sn'(u;k)}{sn(u;k)}-\frac{k\,sn'(\xi;1/k)}{\xi\left(sn(u;k)^2\xi^2-1\right)}\right)\varphi}\\
\qquad \qquad \displaystyle{-\frac{\sqrt{u_x^2+1}}{2\sqrt{sn(u;k)^2\xi^2-1}}\left(\frac{sn'(u;k)}{sn(u;k)}-\frac{k\,sn'(\xi;1/k)}{\xi}\right)\psi, }\\ \\
\displaystyle{\psi_x=\frac{\sqrt{u_x^2+1}}{2\sqrt{sn(u;k)^2\xi^2-1}}\left(\frac{sn'(u;k)}{sn(u;k)}+\frac{k\,sn'(\xi;1/k)}{\xi}\right)\varphi}\\
\qquad \qquad \displaystyle{-\frac{u_x}{2}\left(\frac{sn'(u;k)}{sn(u;k)}+\frac{k\,sn'(\xi;1/k)}{\xi\left(sn(u;k)^2\xi^2-1\right)}\right)\psi,} 
\end{array}\right.\\ \\
\fl\quad\left\{\begin{array}{l}
\displaystyle{\varphi_y=-\frac{u_y}{2}\left(\frac{sn'(u;k)}{sn(u;k)}-\frac{\xi k\,sn'(\xi;1/k)sn(u;k)^2}{\left(sn(u;k)^2\xi^2-1\right)}\right)\varphi-\frac{\sqrt{u_y^2+q}\sqrt{sn(u;k)^2\xi^2-1}}{2sn(u;k)}\psi,}\\ \\
\displaystyle{\psi_y=\frac{\sqrt{u_y^2+q}\sqrt{sn(u;k)^2\xi^2-1}}{2sn(u;k)}\varphi-\frac{u_y}{2}\left(\frac{sn'(u;k)}{sn(u;k)}+\frac{\xi k\,sn'(\xi;1/k)sn(u;k)^2}{\left(sn(u;k)^2\xi^2-1\right)}\right)\psi.}
\end{array}\right.
\end{eqnarray*}

\bigskip

\textbf{Acknowledgments.} The authors express deep gratitude to I.T. Habibullin for useful discussions and constant attention to the work.

\end{document}